\documentclass[aps,prl,twocolumn,showpacs,amsfonts]{revtex4-1}

\usepackage{amsmath, amssymb}
\usepackage{amsfonts}
\usepackage{graphicx}
\usepackage{bm}
\usepackage{color}

\newcommand{\C}[1]{{\mathcal{#1}}}

\begin{document}
\title{Nonequilibrium thermodynamics and glassy rheology}

\author{Eran Bouchbinder$^1$ and J.S. Langer$^2$}
\affiliation{$^1$Chemical Physics Department, Weizmann Institute of Science, Rehovot 76100, Israel\\$^2$Department of Physics, University of California, Santa Barbara, CA  93106-9530}


\begin{abstract}
Mechanically driven glassy systems and complex fluids exhibit a wealth of rheological behaviors that call for theoretical understanding and predictive modeling. A distinct feature of these nonequilibrium systems is their dynamically evolving state of structural disorder, which determines their rheological responses. Here we highlight a recently developed nonequilibrium thermodynamic framework in which the structural state is characterized by an evolving effective disorder temperature that may differ from the ordinary thermal temperature. The specific properties of each physical system of interest are described by a small set of coarse-grained internal state variables and their associated energies and entropies. The dynamics of the internal variables, together with the flow of energy and entropy between the different parts of the driven system, determine  continuum-level rheological constitutive laws. We conclude with brief descriptions of several successful applications of this framework.
\end{abstract}

\maketitle

\section{Introduction}
\label{intro}
\vspace{-0.25cm}

The rheology of glassy solids and dense complex fluids poses intriguing questions of both fundamental and practical importance. Systems of interest include noncrystalline solids such as structural glasses and polymers, soft particulate systems such as emulsions and colloidal suspensions, granular materials, and a wide variety of biological substances. When driven externally, such systems exhibit complex rheological responses that go beyond those of simple fluids and elastic solids. In \cite{BMG_review_2007, Forterre2008, Ovarlez_etal2009, Derkach2009, Verdier2009, Trexler2010, Chen_etal2010, Weeks2012}, we refer the reader to a few recent review papers that we have found helpful in exploring these topics.  Our  theoretical challenge is to describe the complex microscopic dynamics of such systems by relatively simple, macroscopic, continuum equations.

We argue here that there are two related, fundamental aspects of these systems that are essential ingredients of predictive theories. First, there is a natural distinction between slow configurational (i.e. structural) degrees of freedom and fast kinetic degrees of freedom \cite{Berthier2000, Kurchan2005}. This distinction allows us to apply the principles of statistical thermodynamics to the configurational and kinetic subsystems separately, and thus to describe what happens when they fall out of equilibrium with each other in response to external driving forces. Second, we assert that the memory of prior, irreversible deformations of these systems must be carried by properly defined internal state variables, which satisfy physically motivated equations of motion, and whose present values determine subsequent behavior \cite{Coleman1967, Rice1971}. In what follows, we outline this nonequilibrium thermodynamic framework and describe some of its applications.

\section{Time scales, separable subsystems and the laws of thermodynamics}
\label{timescales}

Equilibrium statistical thermodynamics is a well established bridge between microscopic and macroscopic descriptions of physical systems \cite{Kittel1980}. Driven glassy systems, however, are not in thermodynamic equilibrium; thus it sometimes is argued that thermodynamics is not relevant to them. We disagree. The starting point of our counter argument is the observation that the evolution of the configurational (structural) degrees of freedom of a glass is much slower than, and is only weakly coupled to, the fast dynamics of the kinetic-vibrational degrees of freedom.  The configurational degrees of freedom are the mechanically stable positions of the glassy elements  (atoms, molecules, colloidal particles, etc.) in their inherent structures \cite{Stillinger1995, Sciortino1999}. The kinetic-vibrational degrees of freedom are the momenta and the positional fluctuations about the inherent structures, caused by ordinary thermal or Brownian motions.

For the dense systems of interest here, spontaneous configurational rearrangements -- activated transitions from one inherent structure to another -- are extremely infrequent on the time scales of the kinetic-vibrational fluctuations.  It is in this sense that the configurational degrees of freedom are weakly coupled to the fast kinetic-vibrational degrees of freedom. This separation of time scales, and the associated weak coupling, suggests that the two sets of degrees of freedom can be described approximately as two thermodynamic subsystems in weak contact with each other, in analogy to the standard thermodynamic situation in which heat flows between neighboring subsystems.  Our case is different only in the sense that our two subsystems are not spatially separated from each other.

We further assume that the kinetic-vibrational degrees of freedom equilibrate with a thermal reservoir on  microscopic time scales, independent of the driving forces. (Note that, in athermal glassy systems -- e.g. granular materials or foams -- where the ``atoms'' are macroscopic, no such equilibration takes place.) We also assume that the external driving forces act primarily, but not always exclusively, on the configurational degrees of freedom.

To formulate this physical picture more accurately, we need first to write down the laws of thermodynamics for the two subsystems. For simplicity, we specialize to a spatially homogeneous system of volume $V$ under the application of a purely deviatoric (shear-like, non-hydrostatic) stress $\sigma$. Therefore, the first and second laws of thermodynamics for the system as a whole take the form
\begin{equation}
\label{thermo_laws}
\dot{U}_{tot} = V \sigma D_{tot} \quad\hbox{and}\quad \dot{S}_{tot} \ge 0 \ ,
\end{equation}
where $U_{tot}$ is the total internal energy, $D_{tot}$ is the total rate of deformation (strain rate) and $S_{tot}$ is the total entropy. The essence of the two-subsystem idea is that we can assign well-defined internal energies and entropies to both the configurational (denoted by a c-subscript) and kinetic-vibrational (denoted by a k-subscript) subsystems, leading to \cite{Sciortino1999, BLII2009}
\begin{equation}
\label{decomposition}
U_{tot} \simeq U_c + U_k \quad\hbox{and}\quad S_{tot} \simeq  S_c + S_k \ .
\end{equation}
Note that the reservoir degrees of freedom are included in the k-subsystem, which implies that the k-subsystem equilibrates instantaneously with the reservoir. This assumption can easily be relaxed \cite{BLII2009}.
Moreover, note that it is assumed here that we can define entropies in nonequilibrium situations. (See below.)

To make an explicit statement of the first law of thermodynamics for each of the subsystems, assume that the elastic strain is small enough that the total rate of deformation (common to both subsystems) can be decomposed as $D_{tot}\!=\!D_{el}\!+\!D_{pl}$. Here $D_{el}$ is the elastic rate of deformation that involves no configurational changes and no dissipation in the c-subsystem, and $D_{pl}$ is the plastic (inelastic) rate of deformation that involves configurational changes and dissipation. Furthermore, the stress can be decomposed into partial stresses, $\sigma\!=\!\sigma_c\!+\!\sigma_k$, where $\sigma_k$ is a dissipative stress that is associated with the k-subsystem (e.g. a hydrodynamic viscous stress due to the solvent in colloidal glasses). With these definitions, the first law of thermodynamics takes the form
\begin{eqnarray}
\label{subsystems_1st_law_c}
\dot{U}_c &=& V \sigma_c D_{el} + V \sigma_c D_{pl} - Q_{ck} \ ,\\
\label{subsystems_1st_law_k}
\dot{U}_k &=& V \sigma_k D_{tot} + Q_{ck} \ ,
\end{eqnarray}
where $Q_{ck}$ is the rate of heat flow from the c-subsystem to the k-subsystem. Equations (\ref{subsystems_1st_law_c}) and (\ref{subsystems_1st_law_k}) sum to give the standard first law in Eq. (\ref{thermo_laws}).

In this version of the theory, elastic deformations are exclusively associated with the c-subsystem. (This is not the case when hydrostatic stresses are considered \cite{BLI2009, BLII2009, BLIII2009}.) The plastic part of the mechanical power, $V \sigma_c D_{pl}$, is associated with the c-subsystem, which implies that heat exchange between the subsystems is possible only through $Q_{ck}$, which, by itself, might depend on $D_{pl}$. In principle, we  might consider the possibility that part of the plastic power is directly converted into ordinary heat and flows to the k-subsystem, in which case a fraction of $V \sigma_c D_{pl}$ would appear in the first-law equation for the k-subsystem, and only its complementary part in the c-subsystem  \cite{SollichCates2012, Kamrin2013}.

\section{Internal variables and effective temperature dynamics}
\label{internal_variables}

To further examine the physical implications of Eqs. (\ref{subsystems_1st_law_c})-(\ref{subsystems_1st_law_k}) and the second law of thermodynamics, $\dot{S}_c\!+\!\dot{S}_k\!\ge\!0$, we must be more explicit about the functional dependences of the internal energies $U_c$ and $U_k$.  Specifically, we now must introduce internal state variables.

A theory of irreversible deformation and flow must include equations of motion for a set of coarse grained, internal state variables, denoted here by $\{\Lambda_\alpha\}$, where the subscript $\alpha$ is a discrete index that denotes the members of the set \cite{Coleman1967, Rice1971, Ottinger_Book}. For example, this set may contain the numbers of vacancies or flow defects, or the populations of chemical species.  It must be complete enough that knowledge of the current values of the $\{\Lambda_\alpha\}$ suffices to predict future behavior. The $\{\Lambda_\alpha\}$ must, at least in principle, be observable quantities.  For example, the accumulated plastic strain is not an acceptable member of this set because the undeformed state from which it might be  measured is not observable.

In equilibrium, the values $\{\Lambda_\alpha\}\!=\!\{\Lambda^{eq}_\alpha\}$ are determined thermodynamically; hence, internal state variables do not appear explicitly in equations of state. They are essential, however, for describing nonequilibrium dynamics. For simplicity, assume that the $\{\Lambda_\alpha\}$ belong only to the configurational subsystem.  Also assume, as noted above, that  the elastic deformations $E_{el}$ are associated only with the c-subsystem, where $D_{el}\!=\!\dot{E}_{el}$. Therefore, the functional dependences of the internal energies are
\begin{equation}
U_c(S_c, E_{el}, \{\Lambda_\alpha\})\quad\hbox{and}\quad U_k(S_k) \ .
\end{equation}

Statistical mechanics restricts the definition and use of internal state variables in nonequilibrium situations -- a fact that sometimes is missed in the literature. To see this, it is instructive to invert $U_c(S_c, E_{el}, \{\Lambda_\alpha\})$ in favor of $S_c(U_c, E_{el}, \{\Lambda_\alpha\})$. Away from equilibrium, when the $\{\Lambda_\alpha\}$ are not determined by $U_c$ and $E_{el}$, the  entropy $S_c$ must be defined as the logarithm of the number of configurations available at given $U_c$ and $E_{el}$, {\em further constrained} by fixing the nonequilibrium values of the $\{\Lambda_\alpha\}$ \cite{Ottinger_Book, BLI2009, BLII2009}. For this prescription to make sense, however, the equilibrated, constrained entropy $S_c(U_c, E_{el}, \{\Lambda^{eq}_\alpha\})$ must be the same as the equilibrated unconstrained entropy $S_c(U_c, E_{el})$. This can be true only in the thermodynamic limit of indefinitely large systems, and then only if the set $\{\Lambda_\alpha\}$ contains just a small, non-extensive, number of variables. Most importantly, the entropies associated with these variables must be included explicitly in $S_c(U_c, E_{el}, \{\Lambda_\alpha\})$. For example, if one of the $\Lambda_{\alpha}$ is a number of defects, say $N_d$, then an explicit part of $S_c$ must be the logarithm of the number of ways in which those $N_d$ defects can be distributed in the volume $V$ \cite{BLI2009, BLII2009}.

The thermodynamic derivatives of $U_c$ and $U_k$ play central roles in this thermodynamic framework. The configurational stress $\sigma_c$ is given by $V\sigma_c\!=\!(\partial U_c/\partial E^{el})_{S_c,\{\Lambda_\alpha\}}$. Most importantly, there are {\it two} relevant temperatures \cite{Nieuwenhuizen1998, Sciortino1999, BLII2009}:
\begin{equation}
\theta = \left(\frac{\partial U_k}{\partial S_k}\right)\quad\hbox{and}\quad \chi = \left(\frac{\partial U_c}{\partial S_c}\right)_{E_{el}, \{\Lambda_\alpha\}} \ .
\end{equation}
Here, $\theta$ is the ordinary temperature of the kinetic-vibrational subsystem, which is the same as the temperature of the heat reservoir. $\chi$ is the thermodynamic temperature of the configurational subsystem, usually called the ``effective temperature'' \cite{Cugliandolo1997, Nieuwenhuizen1998, Berthier2000, Ottinger2006, Teff_review_Leuzzi, Teff_review_2011}. It characterizes the state of structural disorder of a glassy system, and thus evolves during deformation and flow. Typically, but not always, the configurational degrees of freedom of a glassy system are ``hotter'' than the kinetic-vibrational ones, $\chi\!>\!\theta$.

With these definitions, we can rewrite the second law in the form
\begin{eqnarray}
\label{entropy_product}
\frac{{\C W}(S_c, \{\Lambda_\alpha\})}{\chi} + \frac{V \sigma_k D^{tot}}{\theta} + \left(\frac{1}{\theta}-\frac{1}{\chi} \right) Q_{ck} \ge 0 \ ,
\end{eqnarray}
where
\begin{equation}
{\C W}=V \sigma_c D_{pl}-\sum_{\alpha}\left({\partial U_c\over\partial \Lambda_\alpha}\right)_{S_c, E_{el}} \dot{\Lambda}_\alpha,
\end{equation}
i.e. the difference between rates of plastic work done and energy stored, is the configurational heat produced during deformation and flow. The entropy production terms in the second law inequality of (\ref{entropy_product}) have distinct physical meanings. The first two terms describe the entropy generated due to dissipation in the c- and k-subsystems, respectively.  The last term describes entropy generation due to heat exchange between the two subsystems when $\chi\!\ne\!\theta$.

Because the various terms in the inequality (\ref{entropy_product}) correspond to different -- and putatively independent -- physical processes, one usually adopts a stronger set of separate inequalities, ensuring the non-negativity of each term separately \cite{ColemanNoll1963}. We can ensure the non-negativity of the second term by setting $\sigma_k\!=\!\eta_k D_{tot}$, with a viscosity $\eta_k\!\ge\!0$, which is a standard viscous relation. Note that if $\eta_k\!>\!0$, deformation that is elastic from the perspective of the c-subsystem (i.e. one that does not involve configurational changes), $D_{tot}\!=\!D_{el}$, will produce dissipation in the k-subsystem due to the action of the viscous/dissipative stress $\sigma_k$. The non-negativity of the last term in (\ref{entropy_product}) can be ensured by a Fourier-like relation $Q_{ck}\!=\!A (\chi\!-\!\theta)$, with a non-negative heat transfer coefficient $A$. The remaining inequality, ${\C W}\!\ge\!0$, imposes constraints on any rheological constitutive law.

The final step in the thermodynamic analysis is to use Eqs. (\ref{subsystems_1st_law_c})-(\ref{subsystems_1st_law_k}) to derive heat equations for the evolution of $\chi$ and $\theta$. Using $\chi\dot{S}_c\!\approx\!C_c\dot{\chi}$ and $\theta\dot{S}_k\!=\! C_k\dot{\theta}$, where $C_c$ and $C_k$ are heat capacities of the c- and k-subsystems respectively, we obtain \cite{BLII2009}
\begin{eqnarray}
\label{heat_c}
C_c ~\dot{\chi} &=& {\C W} - A(\chi-\theta)  \ ,\\
\label{heat_k}
C_k ~\dot{\theta} &=& V \eta_k D_{tot}^2 + A(\chi-\theta) \ ,
\end{eqnarray}
and recall that the second law implies ${\C W}\!\ge\!0$.

Equation (\ref{heat_k}) is an ordinary heat equation with viscous dissipation and heat transfer from the c-subsystem appearing as source terms. If, as indicated earlier, we include a thermal reservoir as part of the k-subsystem, then we can assume that the heat capacity $C_k$ is indefinitely large, and that $\theta$ is the constant temperature of that reservoir.  Otherwise, it is a simple matter to generalize these equations to include heat transfer between the k-subsystem and a separate reservoir at temperature $\theta_R$, and thus to describe variations in $\theta$.

Equation (\ref{heat_c}) is a configurational heat equation for the c-subsystem and is, in fact, an evolution equation for $\chi$. ${\C W}$ is a non-negative source term that tends to enhance glassy disorder due to plastic deformation (``rejuvenation''). The second term on the right-hand-side is typically (i.e. for $\chi\!>\!\theta$) a sink term that tends to reduce glassy disorder (``relaxation'' or ``aging'').  Balancing these two terms raises the possibility of reaching a steady state of disorder during persistent flow. To better understand this, we highlight a special feature of our framework. In standard thermodynamics, coupling coefficients like $A$ in Eqs. (\ref{heat_c})-(\ref{heat_k}) depend on state variables such as the ordinary temperature $\theta$. Here, however, the irreversible flow itself can determine  the coupling between the subsystems, and $A$ may be proportional to the plastic power $V \sigma_c D_{pl}$. Therefore, a steady state $\chi\!\ne\!\theta$ naturally emerges, even in the athermal limit, $\theta\!\to\! 0$. This behavior is widely observed \cite{Teff_review_2011, Berthier2002, Ono2002, Makse2002, GM2005, Colloids2006}.

\section{Constitutive laws}

Our thermodynamic framework culminates with the heat equations (\ref{heat_c})-(\ref{heat_k}), which is as far as thermodynamics can take us. From this point on, we need to invoke physical considerations relevant to specific physical systems and phenomena of interest. That is, our thermodynamic framework must be supplemented by rheological constitutive laws.

The hallmark of such constitutive laws is an expression for the plastic rate of deformation $D_{pl}(\sigma_c, \theta,\chi, \{\Lambda_\alpha\})$. To derive and use such an expression, we must choose the internal variables $\{\Lambda_\alpha\}$, determine their associated energies and entropies, and derive their evolution laws $\dot{\Lambda}_\alpha$.  In what follows, we identify several examples of constitutive laws in which the elements of our thermodynamic framework play essential roles.

Significant progress in understanding deformation and flow of glassy materials has been made in the last few decades \cite{ArgonQuo1979, Srolovitz1981, Srolovitz1983, Falk1998, Mayr2006, Schall2007, STZ_BMG_2008, Delogu2008}. While crystalline solids mainly flow by the propagation of dislocations (lattice scale topological defects \cite{HirthLothe1968}), the lack of long-range translational and orientational order in glassy solids implies the absence of dislocation-mediated crystallographic slip. Instead, glassy flow has been shown to be mediated by stress-driven irreversible rearrangements of localized clusters of deformable  elements. These localized clusters -- termed ``flow defects'' or``shear transformation zones'' (STZ's) -- are the elementary carriers of inelastic deformation in these systems. This general concept has been the basis for the modern development of various mesoscopic models of glassy rheology \cite{Spaepen1977, Argon1979, ArgonShi1983, Sollich1997, Falk1998, Demetriou2006}.

An important rheological model of this kind is known as ``soft glassy rheology'' (SGR) \cite{Sollich1997}. In this model, a glassy material is described as a collection of mesoscopic elements, each of which is characterized by a local strain $l$ and a local energy barrier $E$ that must be overcome during rearrangements. Material disorder implies that different elements are characterized by different $l$'s and $E$'s, and the distribution function $p(E,l)$ plays the role of the $\{\Lambda_\alpha\}$. The effective temperature $\chi$, which is called the ``noise temperature'' in SGR, plays an essential role in activating rearrangements. This model has been useful in explaining a variety of soft glassy rheological phenomena \cite{Fielding2000, Fielding2009}. Recently, it has been reformulated so as to be largely consistent with the thermodynamic framework outlined here \cite{SollichCates2012}. In view of the successes of SGR, it will be important to better understand its relations with other models such as the STZ model, which we discuss next.

In the current, thermodynamic version of the STZ model, shear transformation zones (STZ's) are dilute, two-state, flow defects that make transitions between their internal orientations in response to external stresses \cite{Falk1998, BLPI2007, BLPII2007, Langer2008, Falk2011}. The average rate of these transitions is proportional to the rate of irreversible plastic deformation, $D_{pl}$. Importantly, STZ's have finite lifetimes. They are continually being created and annihilated by thermal fluctuations and by the mechanical noise generated by the STZ transitions themselves. The total STZ density $\Lambda$ and an orientation tensor $\mathbf{m}$ constitute the set of internal variables $\{\Lambda_\alpha\}$. The time evolution of $\Lambda$ and $\bf  m$ is determined by master equations describing both the internal transitions and the annihilations and creations discussed above. Explicit expressions for the entropy $S_z(\Lambda,\bf  m)$ and energy $U_z(\Lambda)$ are associated with the STZ population, and consistency with the second law constraint ${\C W}\!\ge\!0$ is enforced. The steady-state STZ density turns out to be proportional to an effective Boltzmann factor $\Lambda\!\sim\!e^{-e_z/k_B \chi}$, where $e_z$ is the STZ formation energy and $k_B$ is Boltzmann's constant, thus providing a direct link between the nonequilibrium thermodynamic framework and the constitutive model \cite{Langer2004, BLII2009, Falk2011}.

\begin{figure}[h]
\centering
  \includegraphics[height=7.5cm]{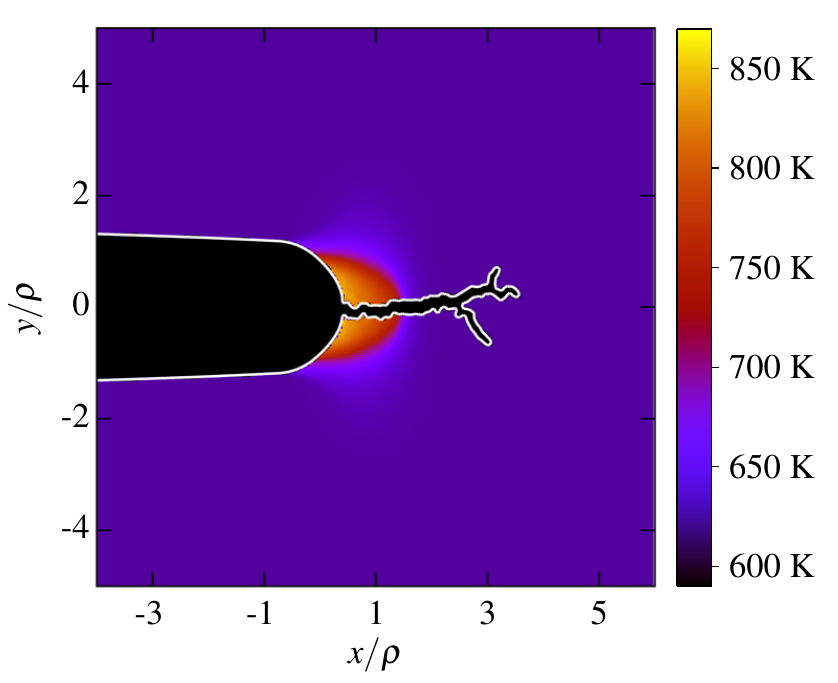}
  \caption{A snapshot of a crack initiation process from an initial notch of radius $\rho$, simulated using the thermodynamic STZ model \cite{Rycroft2012}. The effective temperature field $\chi(x,y)\!>\!\theta\!=\!400$K is plotted, exhibiting strong localization near the notch root, where failure initiates. See additional details in \cite{Rycroft2012}.}
  \label{fig1}
\end{figure}

The STZ model and its variants have been shown to account for a wide range of glassy rheological behaviors, including predictions of stress as a function of strain and strain rate, the appearance of yield stresses at low temperatures, shear banding instabilities and linear viscoelasticity \cite{BLPI2007, BLPII2007, Langer2008, Falk2011, STZ_ShearBanding_2007, STZ_cavitation_2008, STZ_ShearBanding_2009, LinearResponse_PRL_2011, LinearResponse_PRE_2011, Gibou2012}. A notable recent example \cite{Rycroft2012} is the prediction of an annealing-induced brittle-to-ductile transition in metallic glasses (an important class of new materials \cite{Schroers2013}). This calculation uses a space- and time-dependent, tensorial version of the STZ constitutive law, combined with equations of motion for the elastic field and the effective temperature $\chi$ in the neighborhood of a crack tip. Realistic system parameters deduced from earlier analyses of spatially uniform deformation are used \cite{Langer2008}. If the initial $\chi$ is small, the tip sharpens and emits cracks under the influence of an applied stress; but, if the initial $\chi$ exceeds a critical value, the tip continuously blunts and fracture is significantly delayed \cite{Rycroft2012}. These results agree with experiments \cite{Lewandowski2001, Lewandowski2005, Kumar2013}. We emphasize that this calculation uses all of the elements of our thermodynamic framework in essential ways -- especially the dynamics of $\chi$, which couples to the deformation rate and guides shape changes near the crack tip. A snapshot of the failure process near the tip of the crack is presented in Fig. \ref{fig1}.

Two, somewhat different applications of this nonequilibrium thermodynamic framework should be mentioned, at least briefly. First, we have made an analysis of the Kovacs memory effect, in which the volume of a glassy material under constant pressure is measured during a sequence of abrupt temperature changes near the glass transition. Typically, this volume undergoes a sequence of complex, non-exponential, relaxations toward equilibrium, accompanied by marked history dependences. In this case, the relevant internal variables are the populations of vacancies that are created and annihilated in response to the temperature changes \cite{Kovacs_Teff_2010}. Second, and even further afield, extensions of the present ideas to the case of dislocation-mediated plasticity in polycrystalline solids, where the areal density of dislocations is the relevant internal variable, have begun to emerge \cite{Berdichevsky2008, Dislocations_Teff, Berdichevsky2012}. An effective temperature $\chi$ plays a central role in both of these applications.

\section{Acknowledgments}

E.B. acknowledges support from the Minerva Foundation with funding from the
Federal German Ministry for Education and Research, the Israel Science Foundation (Grant No. 712/12), the Harold Perlman Family Foundation and the William Z. and Eda Bess Novick Young Scientist Fund.  J.S.L. was supported in part by the U.S. Department of Energy, Office of Basic Energy Sciences, Materials Science and Engineering Division, DE-AC05-00OR-22725, through a subcontract from Oak Ridge National Laboratory.


\providecommand*{\mcitethebibliography}{\thebibliography}
\csname @ifundefined\endcsname{endmcitethebibliography}
{\let\endmcitethebibliography\endthebibliography}{}

\end{document}